\documentclass[12pt]{article}
\usepackage{cite,graphicx,amsmath,amssymb}
\usepackage{graphicx}
\usepackage{multicol}
\usepackage{xcolor}

\newcommand{\be}{\begin{equation}}
\newcommand{\ee}{\end{equation}}
\newcommand{\bea}{\begin{eqnarray}}
\newcommand{\eea}{\end{eqnarray}}
\begin{document}

\thispagestyle{empty}
\vspace*{.2cm}
\noindent

\vspace*{2.0cm}

\begin{center}
{\Large\bf A Shift Symmetry in the Higgs Sector:\\[.3cm]
Experimental Hints and Stringy Realizations}
\\[2.5cm]

{\large Arthur Hebecker, Alexander K.~Knochel and Timo 
Weigand\\[6mm]}

{\it
Institut f\"ur Theoretische Physik, Universit\"at Heidelberg, 
Philosophenweg 19,\\ D-69120 Heidelberg, Germany\\[3mm]

{\small\tt (\,a.hebecker} {\small ,} 
{\small\tt \,a.k.knochel} {\small and} 
{\small\tt \,t.weigand@thphys.uni-heidelberg.de}\small\tt \,) }
\\[2.0cm]

{\bf Abstract}
\end{center} 

\noindent
We interpret reported hints of a Standard Model Higgs boson at $\sim 125$~GeV
in terms of high-scale supersymmetry breaking with a shift symmetry in the
Higgs sector. More specifically, the Higgs mass range suggested by recent LHC
data extrapolates, within the (non-supersymmetric) Standard Model, to a
vanishing quartic Higgs coupling at a UV scale between $10^6$ and $10^{18}$
GeV. Such a small value of $\lambda$ can be understood in terms of models with
high-scale SUSY breaking if the K\"ahler potential possesses a shift symmetry,
i.e., if it depends on $H_u$ and $H_d$ only in the combination
$(H_u+\overline{H}_d)$. This symmetry is known to arise rather naturally in
certain heterotic compactifications. We suggest that such a structure of the
Higgs K\"ahler potential is common in a wider class of string constructions,
including  intersecting D7- and D6-brane models and their extensions to
F-theory or M-theory.  The latest LHC data may thus be interpreted as hinting
to a particular class of compactifications which possess this shift symmetry. 

\newpage
\section{Introduction and Summary}
While the LHC has so far not produced any significant hint of low-scale
supersymmetry, signals of a Standard Model (SM) Higgs boson at $\sim 125$~GeV have
been reported \cite{exp}. It is clearly far too early to give up on TeV-scale 
SUSY (see e.g. \cite{Papucci:2011wy}). Nevertheless, it may be worthwhile 
investigating what the latest data, if substantiated, imply for string
model building without low-energy supersymmetry. Even if low-energy
supersymmetry were not to be found at the LHC, it is out of question that
string theory remains the most successful candidate for an ultra-violet
completion of gauge and gravitational interactions.  Let us therefore search
for possible interpretations of the announcements \cite{exp} within string 
theory without assuming an imminent discovery of supersymmetry.\footnote{
For 
an alternative approach see e.g. the recent analysis of 
\cite{Aparicio:2012iw}.}

\begin{figure}
\begin{center}
\includegraphics[width=7.8cm]{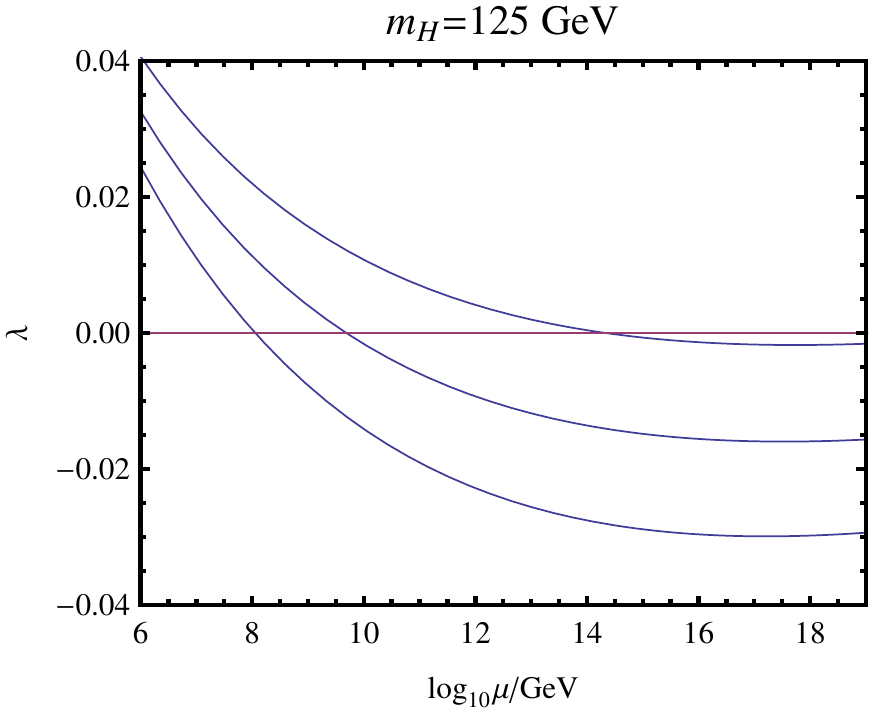}
\includegraphics[width=7.8cm]{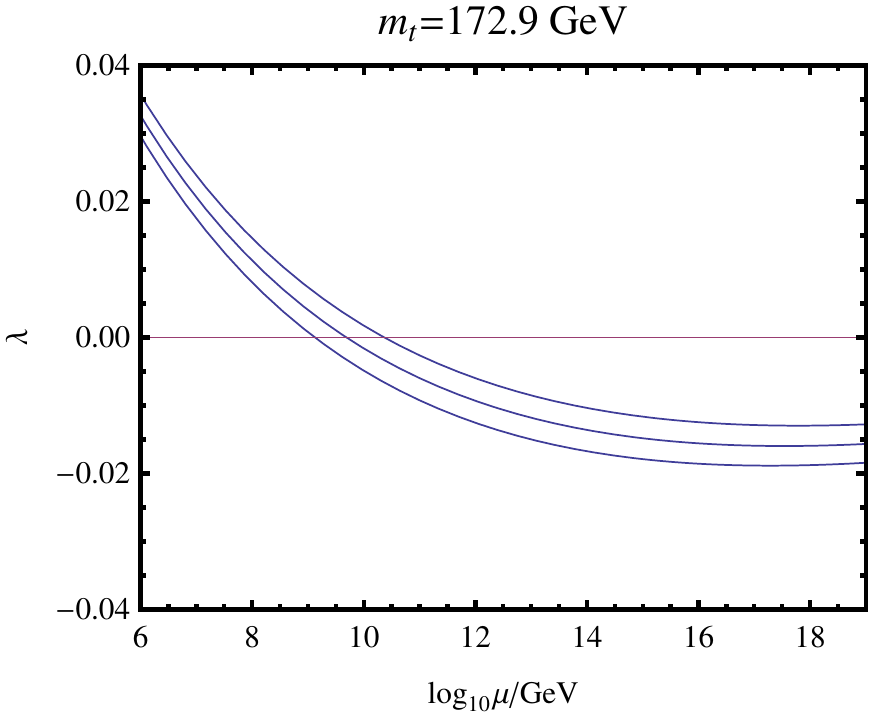}
\end{center}
\caption{The two-loop RG running of the Higgs quartic coupling in the SM.
Left: $m_H=125$ GeV, $m_t=170.7, 172.9, 175$ GeV from top to bottom. Right: $m_t=172.9$ GeV, $m_H=126,125,124$ GeV from top to bottom.
\label{quarticrge}
}
\end{figure}

It has been known for a long time \cite{Lindner:1988ww,Gogoladze:2007qm,
Isidori:2007vm,Shaposhnikov:2009pv} that, within the SM, certain
values of the Higgs mass relate to a vanishing quartic coupling $\lambda$ at
some high but sub-Planckian energy scale. This includes some  
predictions of a $\sim 125$~GeV Higgs \cite{Gogoladze:2007qm,
Shaposhnikov:2009pv} and
has been further discussed in a number of recent papers \cite{
Holthausen:2011aa,EliasMiro:2011aa}. We illustrate the situation in figure
\ref{quarticrge}. In the following, we will be concerned with possible 
origins of the apparently favorable high-scale boundary condition $\lambda=0$. 

The necessary fine-tuning of the electroweak scale within the 
non-supersymmetric SM can arise for example within the flux-based 
string theory landscape (see e.g. \cite{Denef:2004ze}). The best-understood
underlying string compactifications are nonetheless supersymmetric at least 
near the string scale. Hence we assume that at some high energy scale the 
SM is embedded in a supersymmetric theory. This puts us in the 
context of \emph{high-scale} SUSY breaking (see e.g. \cite{ ArkaniHamed:2004fb,
Hall:2009nd,highscalerecent}), where we can start from the tree-level expression
\be
\lambda(m_S)=\frac{g^2(m_S)+g'^2(m_S)}{8}\cos^2 2\beta
\label{effectivelambda}
\ee
for the SM quartic coupling at the SUSY breaking scale $m_S$ 
with $g$ and $g'$ the gauge couplings of $SU(2)_L$ and $U(1)_Y$, respectively. 
The desire to have vanishing $\lambda$ at some high energy scale thus points
towards models with $\tan\beta\simeq 1$. The interesting question is now which
structure of the high-scale model may be responsible for this particular value
of $\beta$. 

A possible answer can be based on a rather old observation in the heterotic
orbifold context \cite{LopesCardoso:1994is,Antoniadis:1994hg,Brignole:1995fb,
Brignole:1997dp} (also investigated more recently in the orbifold GUT context 
\cite{Choi:2003kq,Hebecker:2008rk,Brummer:2009ug,Lee:2011dya}). The 
observation is that, for a certain class of models, the Higgs-sector K\"ahler 
potential possesses a shift symmetry, $H_u\to H_u+c$, $H_d\to 
H_d-\overline{c}$, at tree-level. It hence reads 
\bea \label{Kahler1}
K_H=K_H(H_u+\overline{H}_d,\overline{H}_u+H_d,S,\overline{S})
=|H_u+\overline{H}_d|^2f(S,\overline{S})+\cdots\,,
\eea
where $S$ stands for an appropriate set of moduli.  In section
\ref{sec_Pheno} we will show that this structure indeed corresponds to
$\lambda=0$ at the supersymmetry breaking scale.  We will furthermore analyse
the correlation between the  value of the Higgs mass  suggested by
\cite{exp} and the supersymmetry breaking scale $m_S$ under the
assumption of this shift symmetry. Interpreted as a prediction for $m_S$ in 
terms of the Higgs mass, the concrete result strongly depends on the precise 
value of the top-quark mass and $\alpha_s$. Note that this uncertainty affects 
not only our approach, but is in fact common to all predictions of the Higgs 
mass on the basis of RG running arguments, including e.g. the analyses of 
\cite{Gogoladze:2007qm,Shaposhnikov:2009pv}. We also quantify loop corrections 
to the shift-symmetric tree-level K\"ahler potential from the Yukawa 
couplings.  We argue that these loop effects can be treated as a small 
perturbation and estimate their impact on Higgs mass predictions.

Our philosophy in this letter is to take the phenomenological considerations 
of section~\ref{sec_Pheno} as a motivation to investigate,  in section
\ref{sec_Strings}, how a shift symmetry of type (\ref{Kahler1}) follows from
ultra-violet completions of the SM within string theory.  We will
argue that a shift symmetry of the above type can be realised in string models
where the Higgs sector is related to Wilson line moduli of a higher dimensional
gauge theory\footnote{In this sense our proposal can be viewed a stringy/supersymmetric version
 of \cite{Gogoladze:2007qm}.}. As noted already, examples of this type of bulk Higgs fields 
and the associated shift symmetries are known to arise in orbifold
compactifications of the heterotic string \cite{LopesCardoso:1994is,
Antoniadis:1994hg,Brignole:1995fb,Brignole:1997dp}. We will argue that similar 
structures are possible in suitable Type II compactifications with D-branes, 
thereby taking some first steps towards generalising the framework of 
shift-symmetric Higgs sectors beyond the heterotic orbifold context.

As a starting point it is beneficial to recall the geometric or field
theoretic origin of the shift symmetry within the orbifold constructions of
\cite{LopesCardoso:1994is,Antoniadis:1994hg,Brignole:1997dp} as discussed in
\cite{Brummer:2009ug}.  We are interested in situations where both Higgs
doublets come from the untwisted sector, more specifically from gauge field
components associated with the same complex plane. Such fields are known as
continuous Wilson lines \cite{Ibanez:1987xa}. They may be defined in a
manifestly gauge-invariant manner by the holonomy corresponding to a certain
closed loop in the orbifold (obviously not corresponding to
 an element of $H_1$, but impossible to contract because of the orbifold
singularities). One may also think of the physical meaning of these continuous
Wilson lines in terms of the relative orientation in gauge-space of subgroups 
surviving at distant fixed-points (see section 4 of \cite{Hebecker:2004ce}). 
 
Translated into the language of smooth heterotic models \cite{Honecker:2006qz},
such fields will contribute to certain types of bundle moduli.\footnote{In
addition, the bundle moduli also receive contributions from twisted sector
fields in the orbifold limit,  see e.g. \cite{Tatar:2008zj}.} These in turn
should correspond to brane moduli in Type II orientifolds with D-branes. In 
particular, it can be the Wilson line sector counted by $H^1(\Sigma)$ of a 
D-brane wrapping a cycle $\Sigma$ of the compactification space in which 
Higgs fields with a shift symmetry take their origin.

The presence of a shift symmetry in the tree-level K\"ahler potential and its 
absence in the superpotential can be understood in field theoretic terms as 
follows: A non-zero vacuum expectation value (VEV) of a Wilson line cannot be 
detected by any local observer in the original higher-dimensional theory.  
Thus, tree-level dimensional reduction at quadratic order does not see this 
VEV. However, at cubic order (corresponding e.g. to Yukawa couplings coming 
from gauge couplings) the zero modes of matter fields couple to the Wilson 
line and hence to the Higgs. This is clear since a zero-mode is a global 
object and it potentially feels the holonomy along some closed loop on the 
compact space. Thus, both in smooth heterotic models and in other string 
compactifications (e.g. with gauge theories from branes) we expect analogous 
shift symmetries to arise whenever we have a bundle/brane deformation which 
is {\it a pure gauge transformation} in the (sufficiently small) neighbourhood 
of any point in the higher-dimensional compact space. Its physical reality as 
a true deformation must be associated purely with non-local effects, as is the 
case for the famous heterotic continuous Wilson line.  Furthermore, since we 
need only an approximate shift symmetry phenomenologically, we may be 
satisfied with models which fulfill the above requirement only approximately.  
In section \ref{sec_Strings} we will back up these general considerations 
applied to Type II compactifications with D-branes by a well-known conformal 
field theory argument \cite{Wen:1985jz,Kachru:2000ih}.  
 
The non-trivial task is clearly to construct models where (part of) the MSSM
Higgs degrees of freedom are realized in this way, with a sizeable top Yukawa
coupling still present. It is known from heterotic orbifold constructions with
Gauge-Yukawa-unification \cite{Buchmuller:2005jr} that this is possible in
principle.  In section \ref{sec_Strings} we make some preliminary steps towards
generalising this structure to Type II compactifications with D-branes.  While
the appearance of a Higgs sector from Wilson lines in the above sense seems
natural, a detailed investigation of realistic models is an exciting challenge 
left for future work.

\section{Phenomenology of Higgs sector shift symmetry}
\label{sec_Pheno}

We begin our phenomenological analysis by reviewing in more detail how the
high-scale boundary conditions for the SM quartic coupling
arise in the four-dimensional supergravity picture. We then demonstrate that a
shift symmetry of the K\"ahler potential at high scales is a predictive
assumption even in the presence of the top Yukawa coupling which violates it at
the one-loop level.

Our starting point is the K\"ahler potential (\ref{Kahler1}), and we assume
that $W$ contains no term $\sim H_uH_d$ (i.e. the $\mu$ term is generated
solely through the Giudice-Masiero mechanism).  Without loss of generality, we
take $f=1$ in the vacuum.  Upon supersymmetry breaking, soft masses $m_{H_u}^2,
m_{H_d}^2$, the $B \mu$ term and the effective $\mu$ term are generated.  The
resulting Higgs mass matrix (with $B\mu\equiv m_3^2$), defined through
\be
{\cal L}\supset -m_1^2 |H_u|^2-m_2^2|H_d|^2-m_3^2 (H_u\overline{H}_d+
\overline{H}_uH_d)\,,
\ee
is then given by (cf. e.g.~\cite{Brignole:1997dp})
\be
m_1^2=m_2^2=m_3^2=\left|\mu\right|^2+m_{3/2}^2-F^S \overline{F}^{\overline{S}}
(\ln f)_{S\overline{S}}\,,
\label{mi}
\ee
where 
\be
|\mu|^2=\left|m_{3/2}-\overline{F}^{\overline{S}}
f_{\overline{S}}\right|^2\,,\qquad
\overline{F}^{\overline{S}}=e^{K/2}K^{\overline{S}S}D_SW\qquad\mbox{and}\qquad
m_{3/2}=e^{K/2}W\,.
\ee
The generalization to several moduli instead of $S$ is obvious.  The Higgs mass
matrix (\ref{mi}), which owes its special form $m_1^2=m_2^2=m_3^2$ to the
shift-symmetric K\"ahler potential (\ref{Kahler1}), has the peculiar property
that there is one vanishing eigenvalue with the eigenvector 
\be
H_0 =\frac{1}{\sqrt{2}}(H_u-\overline H_d)\,.
\label{masslesshiggs}
\ee
Since we assume that the soft scale is at least several orders of magnitude
above the electroweak scale, we are thus in the decoupling limit where one
SM-like Higgs doublet, the field $H_0$, remains light and provides the SM Higgs
boson and the would-be Goldstone modes for $W^\pm$ and $Z$.  The orthogonal
combination and thus the states corresponding to the heavy and charged Higgs as
well as the pseudoscalar Higgs in the MSSM become heavy at the soft scale.  The
SM Higgs boson $H_0$  as in (\ref{masslesshiggs}) corresponds to a Higgs mixing
angle $\tan\alpha=-1$.  Since the  electroweak symmetry breaking VEV resides
only in the light Higgs boson in the decoupling limit, this also implies
$\tan\beta = \tan(\alpha+\frac{\pi}{2})=1$. As a consequence, the tree-level
quartic coupling for $|H_0|^4$ originating from the D-term potential of the
electroweak gauge theory vanishes at the soft scale $m_S$ according to
(\ref{effectivelambda}).  Furthermore, in this scenario the gauge and Yukawa
couplings to the light Higgs boson have exactly their respective SM values.
This means for example that, unlike in MSSM-like scenarios with large
$\tan\beta$, the top quark has the only $\mathcal O(1)$ Yukawa coupling.
 
Let us say a few more words concerning the phenomenological implications. We
can base ourselves completely on the analyses of e.g.
\cite{Hall:2009nd,EliasMiro:2011aa}, which give an explicit formula for the SM
Higgs mass based on linearized approximations of the RGE solutions. Adapted to
our case of interest, they imply (with all mass scales in GeV)
\be
m_h=125+1.0\left(\log_{10}\frac{m_S}{4\cdot 10^9}\right)+1.8\left(
\frac{m_t-173.2}{0.9}\right)-0.5\left(\frac{\alpha_s(m_Z)-0.1184}{0.0007}
\right)+\delta\,.
\label{mh}
\ee
This simply states that $\tan\beta=1$ (i.e. $\lambda=0$) at the SUSY breaking
scale $m_S=4\cdot 10^9$ GeV implies $m_h=125$ GeV with a certain set of corrections given in
self-explanatory notation. Obviously, this has the potential of leading to a
rather precise prediction of the SUSY breaking scale. We emphasize, however,
that such a prediction requires a significant improvement of the top and Higgs
mass measurements: At present, a $2\sigma$ shift of $m_t$ together with a 126
GeV Higgs mass allows one to move the point where $\lambda$ vanishes all the
way up to the Planck scale \cite{EliasMiro:2011aa}.\footnote{This is not well
reflected by the {\it linearized} expression in Eq.~(\ref{mh}). We numerically
solve the full two-loop RGEs in the following.}

The crucial issue for us is the intrinsic theoretical uncertainty $\delta$ in
Eq.~(\ref{mh}) associated with violations of the shift symmetry.  Let us
estimate how small this uncertainty might become under favorable circumstances:
We take the K\"ahler potential to be exactly shift-symmetric at the
compactification scale $m_C$. However, the shift symmetry is broken by the
Yukawa couplings in the superpotential. This will feed into the K\"ahler
potential at the very least through field-theoretic loops in the energy range
$m_C>E>m_S$. We may hope to be in a setting where $m_S$ and $m_C$ are closely
related, but an exact equality is hard to imagine (or even to properly define).
We thus need to estimate the magnitude of these shift-symmetry-violating 
corrections to the K\"ahler potential in the supersymmetric theory.  

\begin{figure}
\begin{center}
\begin{picture}(250,180)(100,0)
\put(0,0){
\includegraphics[height=5.5cm]{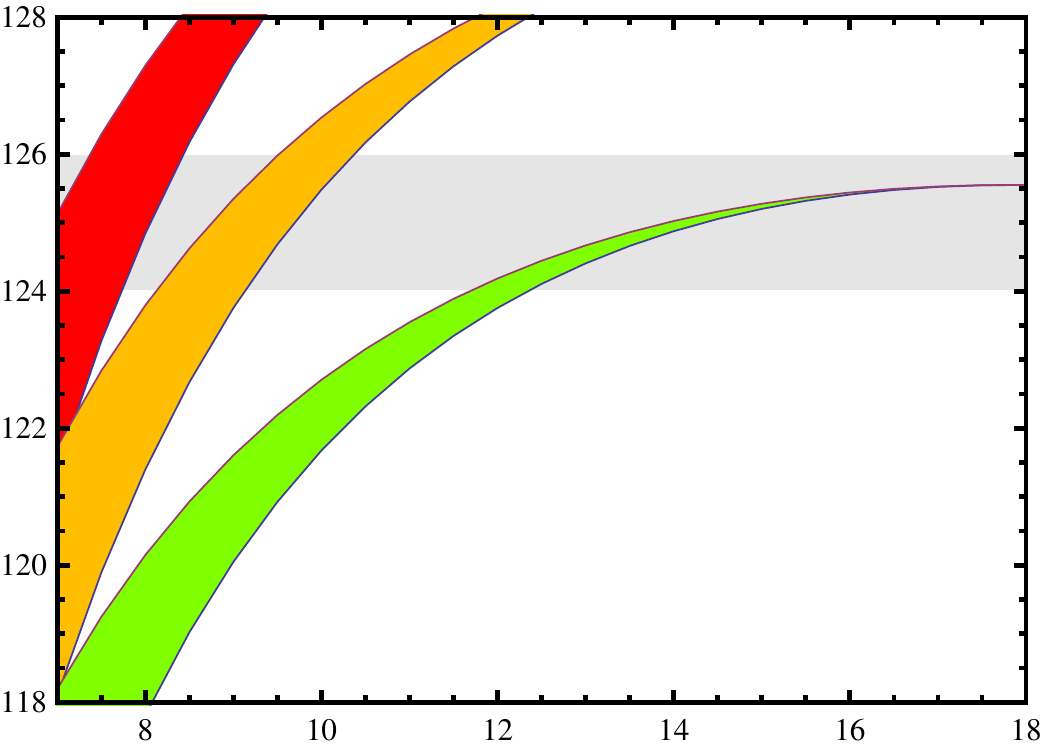}
\includegraphics[height=5.5cm]{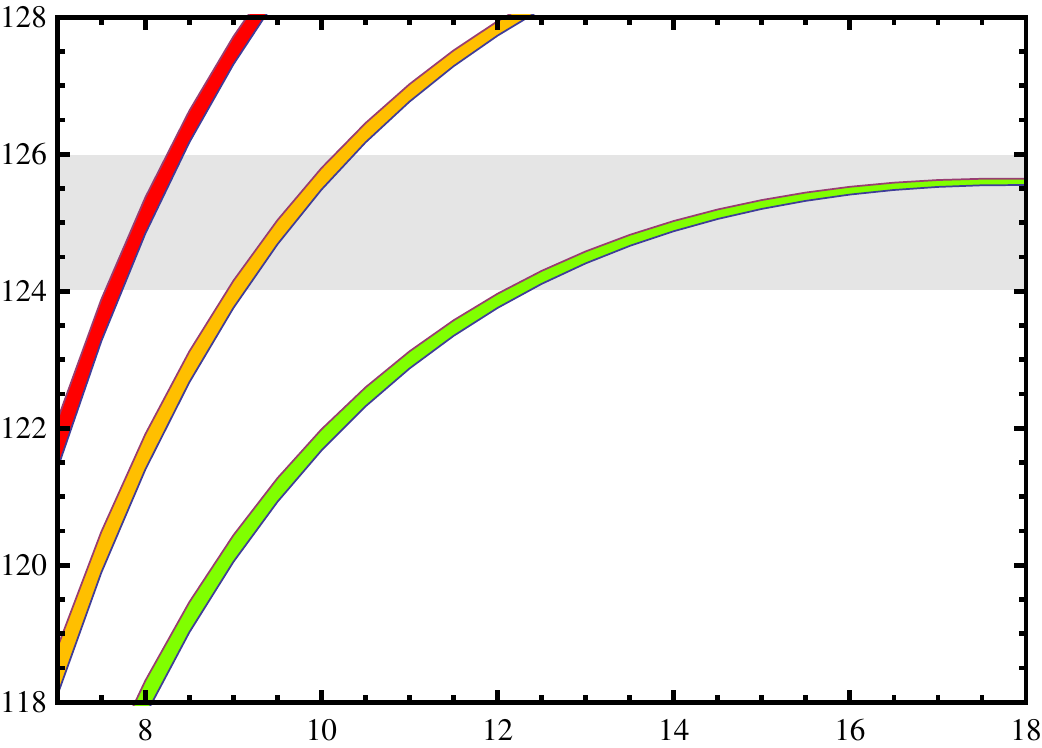}
}
\put(90,-10){$\log_{10} m_S$/GeV}
\put(310,-10){$\log_{10} m_S$/GeV}
\put(-10,60){\rotatebox{90}{$m_h$/GeV}}
\end{picture}
\end{center}
\caption{The Higgs mass as a function of the top mass and the soft breaking
scale for $\cos^2 2\beta\in[0\dots 2\epsilon_y^2]$. Shown here are the case of a
variable compactification scale $m_S=m_C^2/M_{Pl}$ (left) and $m_S=10^{-2} m_C$ (right). The lower (green), middle
(orange) and upper (red) bands correspond to top masses of $m_t=170.7, 172.9$
and $175$ GeV respectively. The strong coupling is fixed at
$\alpha_s(m_Z)=0.1184$. A shaded band $m_h=124\dots 126$ GeV is included for
orientation. \label{higgsmassscan}
}
\end{figure}

We expect the leading contributions to the K\"ahler potential
to be those of the rigid SUSY limit. These are proportional to $H_u\overline H_u$ and $\overline H_d H_d$ from Higgs
self-energy graphs involving gauge and Yukawa couplings.
We can thus write the one-loop contribution as  ($t=\ln Q/Q_0$)
\begin{equation}
\frac{d}{dt} K \sim - 2 \gamma_{H_u}(S,\overline S)\, H_u \overline H_u\,-2 \gamma_{H_d}(S,\overline S)\, \overline H_d H_d\,.
\label{kaehlerscaling}
\end{equation}
The corresponding one-loop anomalous dimensions in the MSSM are
\be
\gamma_{H_u}\Big|_{\theta=\overline\theta=0}\sim \frac{3}{16 \pi^2}\left(|y_t|^2-\frac12 g_2^2-\frac{1}{10} g_1^2\right), \quad
 \gamma_{H_d}\Big|_{\theta=\overline\theta=0}\sim \frac{3}{16\pi^2}\left(-\frac12 g_2^2-\frac{1}{10} g_1^2\right)\,. \ee 
The coupling constants are given by $y_t=\sqrt{2}y_t^{SM}$ for
$\tan\beta=1$, $g_2=g$ and $g_1=\sqrt{5/3}\,g'$ in terms of the usual SM
quantities. 
In the presence of supersymmetry breaking, the resulting running of the soft parameters is captured by the moduli dependence of the anomalous dimensions\cite{Giudice:1997ni,ArkaniHamed:1998kj}.
The latter can be inferred from the moduli dependence of gauge and Yukawa couplings and Z-factors according to
\be
\gamma_{H_u}(S,\overline S)=\frac{3}{16\pi^2} \left(\frac{|y_t(S)|^2}{Z_{Q_3}(S,\overline S) Z_{u_3}(S,\overline S)}-Z_{H_u}(S,\overline S) \left( \frac{g_2^2(S,\overline S)}{2}+ \frac{g_1^2(S,\overline S)}{10}\right)\right)
\ee
and analogously for $\gamma_{H_d}$.
In the rigid SUSY limit, the corrections to the K\"ahler potential (\ref{kaehlerscaling}) give rise to the well-known 
running of soft terms and $\mu$. Thus, we can equivalently work with MSSM one-loop RGEs \cite{Martin:1997ns} 
\begin{align}
&16 \pi^2 \frac{d\mu}{dt}\sim 3\mu \left(|y_t|^2-g_2^2-\frac{1}{5} g_1^2\right), \nonumber \\
&16 \pi^2 \frac{d B\mu}{dt} \sim  3 B\mu \left(|y_t|^2-g_2^2-\frac{1}{5} g_1^2\right)+6 \mu \left(|y_t|^2 A_t  + g_2^2 M_2 + \frac15 g_1^2 M_1 \right)  ,\nonumber \\
&16 \pi^2 \frac{d m_{H_u}^2}{dt} \sim 6 |y_t|^2 \big(m_{H_u}^2 + m_{Q_3}^2 + m_{u^c_3}^2\big)+6|y_t|^2 |A_t|^2-6 g_2^2 |M_2|^2-\frac65 g_1^2 |M_1|^2,\nonumber \\
&16 \pi^2 \frac{d m_{H_d}^2}{dt} \sim -6 g_2^2 |M_2|^2-\frac65 g_1^2|M_1|^2
 \,.
\label{mssmrges}
\end{align}
For $M_1=M_2$, the dependence of these RGEs on the dimensionless parameters 
can be expressed entirely through the functions $\beta_y= 6 |y_t|^2/16\pi^2$ and $\beta_g = 3 (-g_2^2-\frac15 g_1^2)/16 \pi^2$.
It is therefore useful to define 
\be\epsilon_{y,g} \equiv\int_{\ln m_C}^{\ln m_S} \beta_{y,g}(t) dt\,\ee
as the small parameters controlling the corrections.
We now want to estimate the impact of shift-symmetry-violating interactions
on $\tan\beta$ in this setting. 
As discussed above, we evaluate the soft parameters
according to (\ref{mi}) at the scale $m_C$ and use their conventional
renormalization group evolution down to the scale $m_S$ giving us the mass
matrix $m_1^2(m_S)\dots m_3^2(m_S)$. 
The resulting mass matrix is generically nonsingular, and we are (unsurprisingly) confronted with the
Higgs hierarchy problem: a finely tuned contribution to the Higgs
potential (which we expect to be similar in size to the radiative corrections
above) must arise in order to have $v\ll m_S$. The resulting massless state is
then given by 
\be H_0 \sim
\frac{1}{\sqrt{2}} \left(-1\pm\frac12 \cos 2\beta\right) H_u+\frac{1}{\sqrt{2}}\left(1\pm\frac12 \cos 2\beta\right)\overline H_d 
\label{smhiggsvector}
\ee
where the value of $\cos 2\beta\ll 1$ depends on the exact form of the
correction which tunes the electroweak scale small.  
We can now consider two types of corrections: for small hierarchies with
$|\ln m_S/m_C|\lesssim 1$, further contributions to the mass matrix beyond
those considered here can easily arise and tune the electroweak scale small. For
larger hierarchies $m_S\ll m_C$, it appears more consistent
to tune the weak scale entirely within a leading-log approximation: We tune
mass parameters $\mu$, $M_i$, $m_Q^2$, $m_u^2$ and the trilinear coupling
$A_t$ such that the leading-log corrected matrix defined by (\ref{mssmrges}),
\be
M(m_S)=
(|\mu|^2+m_H^2)\left[
\begin{array}{cc}
1 & 1  \\
 1  & 1
\end{array}
\right]
+
\left[
\begin{array}{cc}
\delta|\mu|^2+\delta m_{H_u}^2 & \delta B\mu \\
 \delta B\mu  & \delta|\mu|^2+\delta m_{H_d}^2
\end{array}
\right]\,,\ee
becomes singular. To leading order in $\epsilon$, the condition for vanishing determinant of the scalar mass matrix at the soft scale is
\be2 \delta |\mu|^2+\delta m_{H_u}^2+\delta m_{H_d}^2 - 2 \delta B\mu=0\,,\label{tuningcondition} \ee
which for universal stop, higgs and electroweak gaugino soft masses $m_{\tilde q}^2$, $m_H^2$ and $M_{1/2}$ corresponds to 
\be\left[2 m_{\tilde t}^2+ (A_t - \mu)^2\right] \epsilon_y + 2\left[2 M^2_{1/2} -m_H^2 +2 \mu M_{1/2}  + \mu^2 \right]\epsilon_g=0\,. \ee
Assuming that the condition (\ref{tuningcondition}) is satisfied, 
the corrected mass matrix to first order in $\epsilon_i$ 
 has by assumption one eigenvector with eigenvalue $\mathcal{O}(\epsilon^2)$ which approximately corresponds to 
the light Higgs state. It is given by 
\be
(-1+\frac12 \frac{\delta m_{H_u}^2-\delta m_{H_d}^2}{|\mu|^2 +m_H^2},1)+\mathcal{O}(\epsilon^2)\,
\ee
and by comparing with (\ref{smhiggsvector}) we can directly read off 
\be\cos2\beta=\epsilon_y\,\frac{m_H^2+m_{Q_3}^2+ m_{u_3}^2 +|A_t|^2}{2(|\mu|^2 +m_H^2)} \ee
and thus $\cos^2 2\beta\sim \epsilon_y^2$ up to $\mathcal{O}(1)$ factors.
The effective quartic coupling at the soft breaking scale, up to other SUSY
thresholds \cite{Hall:2009nd}, is then given by 
\be\lambda(m_S)= \frac{1}{8}
\left(g^2(m_S)+g^{\prime 2}(m_S)\right) \cos^2 2\beta\,.\ee 
The impact of these shift-symmetry-violating effects on the Higgs mass is
illustrated in Figure \ref{higgsmassscan} using the SM two-loop RGEs for the
quartic, Yukawa and gauge couplings. For the sake of concreteness, we plot the
cases $m_C^2=m_S M_{Pl}$ and $m_S=10^{-2} m_C$ for $\cos^2 2\beta=0\dots 2 \epsilon_y^2$. 

 We can conclude that the tree level shift symmetry in the K\"ahler potential
has predictive power despite the presence of the top Yukawa coupling as long as
some reasonable assumptions about the nature of SUSY breaking are made. Due to
the large impact of the top Yukawa coupling on the running of the quartic
coupling, our prediction of the soft scale $m_S$ for $m_h= 124\dots 125$ GeV
varies between $m_S\sim 10^6\dots 10^{19}$ GeV. This situation will of course
improve with more precise knowledge of the top quark and Higgs boson mass. The
soft scales and compactification scales which appear in our model for
intermediate values of the top quark mass are in a favourable range for
neutrino mass generation via the seesaw mechanism and, if they are related to
$f_a$, to make the axion a realistic dark matter candidate.
Note that for one of the parameter points satisfying $m_C^2=m_S M_{Pl}$, namely
$m_S\approx 10^9$ GeV and $m_C\approx 10^{14}$ GeV,
the coupling constants of $SU(2)_L$ and $U(1)_Y$ unify at
the compactification scale $m_C$ in standard GUT normalization. This might be
interesting in the context of models with symmetry breaking patterns of the
type discussed in this work. Of course, these relations can vary depending on
the precise embedding of the Higgs field into the adjoint representation of the
relevant gauge group.  

\section{Towards a stringy model}
\label{sec_Strings}

In this section we argue that in certain string models a class of bulk matter
enjoys a leading-order shift symmetry in the low-energy effective action that
gives rise to a Higgs K\"ahler potential of the type (\ref{Kahler1}). For this
to happen the bulk matter must be related to Wilson line moduli, whose typical
shift symmetry is then responsible for the advocated structure. While
originally observed in the context of heterotic orbifold models, this
phenomenon is much more general and includes Type II orientifold models with
bulk matter along D-branes. 

\subsection{Shift symmetry in heterotic orbifold models}
\label{HetShift}

A K\"ahler potential of the structure (\ref{Kahler1}) for the Higgs field
arises in heterotic orbifold models where the Higgs fields emerge from states
in the untwisted sector
\cite{LopesCardoso:1994is,Antoniadis:1994hg,Brignole:1995fb}. Such excitations
are the remnants of the internal polarisation states associated with the
original ten-dimensional $E_8 \times E_8$ (or Spin(32)$/{\mathbb Z_2}$) vector
multiplet after imposing the orbifold projection. They therefore propagate as
"bulk matter" on the internal torus orbifold. By contrast, states from
the so-called twisted sector, which are not inherited from the original vector
multiplet, do in general not exhibit a K\"ahler potential of the form
(\ref{Kahler1}).  Such states are present only after taking the orbifold
quotient, and satisfy twisted boundary conditions. They are localised at the
orbifold fixed-points. 

Specifically, \cite{LopesCardoso:1994is} has analysed ${\cal N}=1$
supersymmetric $(0,2)$ $\mathbb Z_N$ heterotic orbifolds in which the internal
six-torus factorises as $T^6 = T^4 \times T^2$. The K\"ahler potential
associated with moduli of the $T^2$ factor takes the form
\bea \label{Kahler2}
K = - {\rm ln} [ (T + \bar T)(U+\bar U) - (B +\bar C)(\bar B + C)],
\eea
where $T$ and $U$ represent the K\"ahler and complex structure modulus of the
$T^2$.  This was derived in the supergravity analysis of
\cite{LopesCardoso:1994is} as a consequence of the coset structure of the
moduli space characteristic for such models.  The role of the Higgs fields
$H_u$ and $H_d$ in (\ref{Kahler1}) is played by the complex Wilson line moduli
$B$ and $C$ arising as combinations of the the Wilson lines of the gauge field
along the two one-cycles of $T^2$.  These supergravity results agree with the
CFT analysis of  \cite{Dixon:1989fj,Antoniadis:1994hg}, where the K\"ahler
potential of heterotic $(2,2)$ models has been computed to second order in
momenta by explicit evaluation of 4-point scattering amplitudes. Indeed the
above structure of the K\"ahler potential was confirmed, where the role of $B$
and $C$ is played by chiral superfields in representation ${\bf 27}$ and ${\bf
\overline{27}}$ of $E_6$ as long as these emerge from the untwisted sector.
The shift symmetry receives corrections at one-loop order in the string
coupling $g_s$, as computed in \cite{Antoniadis:1994hg}.
Furthermore, there are possibly corrections from couplings to twisted
sector fields encoded in higher order (but at treelevel in $g_s$) $n$-point amplitudes.  As pointed out in
\cite{Antoniadis:1994hg}, if the gauge group is further broken e.g. by discrete
Wilson lines as in the Hosotani-Witten mechanism, the descendents of these
fields continue to exhibit the desired structure in the leading-order
tree-level K\"ahler potential.  Implications of the K\"ahler potential
(\ref{Kahler2}) for generation of a $\mu$-term and related phenomenological
aspects have been analysed in \cite{Brignole:1995fb}.

Interestingly a Higgs sector emerging from untwisted matter in the bulk of the
heterotic orbifold is a characteristic of more recent, realistic heterotic
orbifold model building. This structure offers, among other things, a solution
to the doublet-triplet splitting problem. For constructions of this type and
further references see e.g. \cite{Buchmuller:2005jr}.

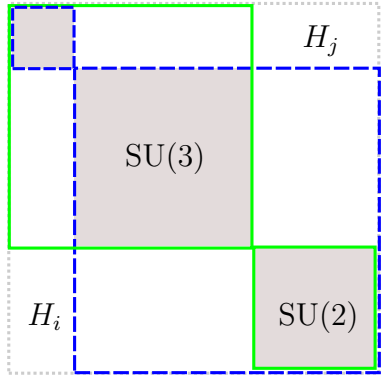
\begin{figure}
\begin{center}
\begin{picture}(150,100)
\put(0,0){\includegraphics[width=5cm]{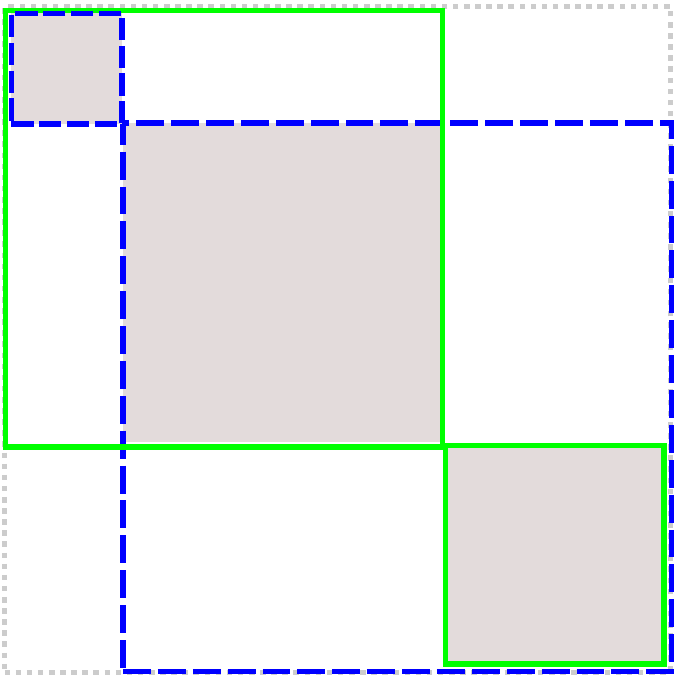}}
\put(8,20){\bf $H_i$}
\put(112,125){\bf $H_j$}
\put(45,80){SU(3)}
\put(103,20){SU(2)}
\end{picture}
\end{center}
\caption{An illustration of the group theoretic origin of Higgs doublets from
Wilson lines in the $SU(6)$ orbifold case.  The components of $su(6)$ are
displayed as a 6$\times$6 matrix. The generators of the gauge groups on the
fixed points, $SU(5)\times U(1)$ and $SU(4)\times SU(2)\times U(1)$, are marked
by a dashed blue and solid green border respectively. The components
corresponding to unbroken generators are shaded. The coset which corresponds to
generators broken on both fixed points (and thus the massless components of
$\Phi$) is marked with $H$.
\label{figcoset}}
\end{figure}

The origin of untwisted matter states as descendents of the ten-dimensional
gauge multiplet suggests an interpretation of the peculiar structure of the
K\"ahler potential as a remnant of the original gauge symmetry. To leading
order, this gauge symmetry results in a shift symmetry for the Wilson line
fields.  As detailed in \cite{Hebecker:2008rk,Brummer:2009ug}, this
interpretation is particularly natural in models which allow for a
five-dimensional limit.  To illustrate the role of the underlying Wilson line
shift symmetry we follow these references and consider a four-dimensional
${\cal N}=1$ supersymmetric $SU(5)$ GUT model that results from a
five-dimensional  $SU(6)$ model compactifed on an $S^1$ orbifold.\footnote{The
assumption of $SU(3)\times SU(2)\times U(1)$ gauge unification is only made for
simplicity, and is not necessary in order to obtain Higgs fields in the correct
representations from Wilson lines. On the contrary, in absence of low-energy
supersymmetry the GUT idea is less compelling.} Prior to orbifolding, the
theory contains a four-dimensional chiral superfield $\Phi$ in the adjoint of
$SU(6)$ whose real and imaginary part are, respectively, the scalar component
of the five-dimensional vector multiplet and the Wilson line along $S^1$.  The
Wilson line shift symmetry, which descends from five-dimensional gauge
invariance, ensures that the tree-level K\"ahler potential to quadratic order
in $\Phi$ takes the form 
\bea \label{Kahler3}
K = \frac{1}{2} \, {\rm tr}(\Phi + \bar \Phi)^2  \, f(S, \bar S)
\eea
with $S$ collectively denoting other moduli fields.  It is possible to choose
the orbifold action in such a way that under the decomposition
\bea
SU(6) \rightarrow SU(5) \times U(1), \qquad  {\bf 35} \rightarrow {\bf 24} + 
{\bf 1} + {\bf 5} + \bar{\bf 5}
\eea
only the component ${\bf 5} + {\bf \bar 5}$ survives for $\Phi$.  A suitable
choice of discrete Wilson lines will lead to a breaking of $SU(6)\rightarrow
SU(4)\times SU(2)\times U(1)$ on one of the fixed points and thus $SU(5)
\rightarrow SU(3) \times SU(2) \times U(1)$. This ensures that $\Phi$ gives
rise precisely to the chiral superfields $H_u$ and $H_d$ in the massless
spectrum. For details of this projection see \cite{Burdman:2002se} and 
\cite{Hebecker:2008rk} for a 6d version and other variants.  The
situation is illustrated in Figure \ref{figcoset}.
In this case evaluation of (\ref{Kahler3}) results in
\bea \label{Kahler4}
K =  \frac{1}{2} \, {\rm tr}(\Phi + \bar \Phi)^2  f(S, \bar S) = (H_u + \overline{H}_d)(\overline{H}_u + {H_d}) \, f(S, \bar S),
\eea
which is the starting point of our phenomenological analysis.

A similar structure of the
K\"ahler potential should be encountered also in smooth heterotic
compactifications and in Type II string constructions with
D-branes.\footnote{As is well known the target space dynamics (D-terms of
anomalous $U(1)$s) drives heterotic orbifold models  away from the orbifold point into the regime of
compactification on at least partially resolved Calabi-Yau manifolds.} However,
a direct comparison between the orbifold point and smooth heterotic
compactifications and furthermore with dual Type II models with branes is
complicated.
The intricate relation between heterotic orbifolds and smooth compactifications
has been explored systematically only in the recent literature  beginning with
\cite{Honecker:2006qz}. As pointed out above, the K\"ahler potential
(\ref{Kahler2}) can be corrected at subleading order (but still at string
tree-level)  by couplings involving pairs of twisted sector blow-up modes.
Giving a non-zero vacuum expectation value to these fields smoothens out the
orbifold into a heterotic compactification with vector bundles on an at least
partially resolved Calabi-Yau space. In the presence of higher order terms
involving these blow-up modes, the shift symmetry might well be broken and the
structure (\ref{Kahler2}) can be corrected.

With this in mind we take the established existence of the desired shift
symmetry in heterotic orbifolds as an  inspiration to search for similar shift
symmetries in Type II models. We will make sure, though, to give arguments
applicable entirely within Type II theory and independent of a possible 
duality to heterotic orbifolds.

\subsection{Shift symmetry for open string Wilson lines}
\label{sec_ShiftOP}

At leading-order the Wilson line moduli $\Phi^{(i)}$
associated with the theory on a (single) Type II D-brane enjoy a 
shift symmetry suitable for our purposes.
For definiteness consider a single D6-brane in Type IIA string theory wrapping
a special Lagrangian 3-cycle $\Sigma$ with $b_1(\Sigma)$ brane moduli. Each
modulus is described by an ${\cal N}=1$ chiral superfield $\Phi^{(i)}$ whose
bosonic component $\varphi^{(i)} + i a^{(i)}$ is the sum of a normal
deformation $\varphi^{(i)}$ and the Wilson line $a^{(i)} = \int_{C_i} A $
(where $C_i$ is one of the $b_1(\Sigma)$ 1-cycles on $\Sigma$). The key 
observation is that in absence
of charged matter states at the intersection of two D6-branes  the Wilson line
$a^{(i)}$ does not couple non-derivatively in the effective action. This holds
at tree-level in $g_s$ and perturbatively in $\alpha'$.  Let us recall the
underlying CFT argument of \cite{Kachru:2000ih}, which generalises similar
arguments from the heterotic string \cite{Wen:1985jz}: Non-derivative couplings
of the Wilson line involve the zero-momentum limit ($k=0$) of the Wilson line
vertex operator. In the  0-picture and in integrated form it is given by
\begin{equation} \label{vertexA}
V_{a^{(i)}}|_{k=0} = \int_{\partial D} A^{(i)}_\mu(X)  \partial_\alpha X^\mu\,d
\sigma^\alpha.
\end{equation}
Here $D$ is the open string worldsheet that appears in the scattering process,
which at tree-level in $g_s$ is topologically a disk, and $A^{(i)}_\mu$ is
polarised parallel to the brane. Since the Wilson line is flat, $dA^{(i)}
=0$, we can write $A^{(i)} = d \chi^{(i)}$ if $X(\partial D)$ is topologically 
trivial in $\Sigma$. Therefore the vertex operator vanishes by integration by 
parts \cite{Kachru:2000ih}. Whenever the above reasoning goes through, no non-derivative terms 
involving the Wilson line are possible in the effective action. By 
holomorphicity this in particular excludes non-derivative terms involving the 
superfield $\Phi^{(i)}$ in the superpotential. More importantly for us, it is 
also the origin of the shift-symmetric form of the brane-modulus K\"ahler 
potential. 

The resulting shift symmetry is corrected at higher order: First, for
topologically non-trivial $X(\partial D)$, as occur for scattering processes
described by a worldsheet instanton, the above argument does not apply. This is
in agreement with the non-perturbative break-down of the classical
shift-symmetry through superpotential terms for A-type brane moduli depending
on $e^{\Phi^{(i)}}$.  Second, if we compute a coupling involving, in addition
to $V_{a^{(i)}}$, boundary changing vertex operators, the disk boundary is
partioned into several segments, and integration by parts need not give zero.
Therefore non-derivative terms in the effective action involving the Wilson
line fields and charged open string states located at the intersection of brane
pairs are possible. This explains in particular the appearance of
superpotential terms of the type $W \supset \Phi^{(i)} \, \cal O$ with $\cal O$
a product of charged localised states. Such terms likewise break the shift
symmetry.  Third, on higher genus worldsheets $V_{a^{(i)}}|_{k=0}$ need not
vanish either, resulting in string-loop corrections in $g_s$ to the effective
action which may involve contact terms in $a^{(i)}$.

Applying the above logic to the K\"ahler potential, we conclude that to second
order in $\Phi^{(i)}$ (and assuming just a single Wilson line for simplicity)
it must exhibit a shift symmetric form of the type $K =( \Phi + \bar \Phi)^2
f(S, \bar S)$, where  $S$ now collectively denotes the closed string moduli of
the compactification.  In addition to this, subleading contributions to the
K\"ahler potential exist which break the shift symmetry. As discussed they are
suppressed by $g_s$, non-perturbative in $\Phi$ and  $\bar \Phi$ or possibly
involve boundary changing charged fields. 
We will comment on the latter in the next section.

Note that these general assertions only depend on the structure of the open
string CFT and must therefore hold independently of the chosen
background.\footnote{A systematic computation of the K\"ahler metric $G_{\Phi
\bar{\Phi}}$ in particular of untwisted matter fields on toroidal backgrounds
has been performed in \cite{Lust:2004cx}, in agreement with results found by
duality with the heterotic string \cite{Aldazabal:1998mr}.} Even though
presented in the context of Type IIA D6-branes, the above worldsheet argument
also governs the structure of Wilson line moduli
on B-type branes, i.e. of D9, D7 or D5-branes in Type IIB orientifolds.

For D7 and D5-branes another set of moduli appears in the form of transverse
deformation moduli. Being transversely polarised as opposed to parallel to the
brane as in (\ref{vertexA}), no direct analogue of the CFT argument for Wilson
lines can be made. This is in agreement with the generic appearance of a
perturbative (in $\alpha'$) superpotential for these deformation modes in
presence of suitable fluxes.  Similarly, it is a priori not clear that the
K\"ahler potential exhibits a shift symmetry (at leading order in a suitable
expansion) because the worldsheet instanton corrections of Type IIA generically
map to perturbative corrections in Type IIB which spoil the shift
symmetry.\footnote{This does not exclude the possible appearance of such shift
symmetries for specific geometries or in particular regions of the moduli 
space \cite{wip}.}

This identifies, at least in the supergravity limit of controllable worldsheet
instanton corrections, the Wilson line moduli sector as a starting point to
achieve a K\"ahler potential of the desired structure.

\subsection{  Shift symmetry for  bulk matter  }

The analysis in the previous section has been for a single D-brane with gauge
group $U(1)$.  We now assess to what extent it is possible to generalise this
to a leading-order shift symmetry  in the K\"ahler potential of Wilson moduli
transforming in the adjoint representation of the gauge group $G$ along a stack
of coincident D-branes.  Upon gauge symmetry breaking such a symmetry induces a
corresponding symmetry for those bulk matter states that descend from $\Phi$,
similarly to the heterotic orbifold summarised in section \ref{HetShift}.
While a more quantitative analysis goes beyond the scope of this letter and is
reserved for future work, we set out to describe the general picture.

Consider first a stack of Type IIA  D6-branes carrying gauge group $G$ and
assume that the 3-cycle $\Sigma$ within the Calabi-Yau 3-fold $X_3$ wrapped by
this stack admits one geometric modulus. The associated chiral superfield
$\Phi$ transforms in the adjoint of $G$.  Under a breaking of the gauge group
$G \rightarrow H \times F$
the adjoint of $G$ decomposes as
\bea \label{ad-dec}
{\rm ad}(G) \rightarrow ({\rm ad}(H),1)  \oplus (1,{\rm ad}(F))  \oplus  
\sum_i [ (R_i,U_i) + c.c. ],
\eea
where $R_i$ and $U_i$ are  irreducible representations of $H$ and $F$,
respectively. Concretely we are interested in set-ups in which the surviving
gauge group $H$ either is directly $SU(2)_L \times U(1)_Y$ or contains it (in
which case another breaking mechanism would have to be implemented in a second
step).  The gauge symmetry breaking must be such that the only surviving
components of $\Phi$ can be identified with $H_u$ and $H_d$. The SM
$SU(3)$ factor can be associated with a different brane stack as is common in
intersecting brane models, or it can be contained within $H$ in a GUT
construction. 

The vertex operator argument presented around eq. (\ref{vertexA}) for a single
D-brane makes use of the fact that the amplitude does not involve boundary
changing operators. In particular, the Wilson line $a^{(i)}$ itself is the
excitation of an open string starting and ending on the same brane.  For a
stack of $N$ coincident D-branes at generic position (i.e. not invariant under
the orientifold projection) the original gauge group is $G = U(N)$. The Wilson
line shift symmetry for $N=1$ as realized by a single brane directly carries
over to a shift symmetry for the components of $\Phi$ associated with 
each of the $N$ Cartan generators $U(1)^N \subset U(N)$. These correspond to 
states from open strings starting and ending on one of the $N$ branes within 
the stack. The shift symmetry of these fields together with the full $U(N)$ 
gauge invariance constrains the possible couplings in a manner sufficient 
for our purposes: 

Indeed, let $\Phi=\Phi^a T^a$, where $T^a$ are the $N^2$ generators of 
$U(N)$. We can restrict our attention to the $N^2-1$ generators 
of $SU(N)\subset U(N)$. Within this subset of fields there is clearly no 
gauge invariant term linear in $\Phi$. At quadratic order, the most general 
K\"ahler potential can be built from the two independent sets of fields 
$(\Phi+\bar{\Phi})^a$ and $(\Phi-\bar{\Phi})^a$, contracted in all possible 
ways with the unique invariant tensor, $\delta^{ab}$. Of the resulting three 
terms 
\be 
(\Phi+\bar{\Phi})^a(\Phi+\bar{\Phi})^b\delta^{ab}\,,\quad
(\Phi+\bar{\Phi})^a(\Phi-\bar{\Phi})^b\delta^{ab}\,,\quad
(\Phi-\bar{\Phi})^a(\Phi-\bar{\Phi})^b\delta^{ab}\,,
\ee
the last two are forbidden for the Cartan generators and hence forbidden 
in general. This ensures the desired shift-symmetric structure of the 
quadratic-order K\"ahler potential.

At cubic order in $\Phi$, one can use the structure constants $f^{abc}$ and 
build invariant expressions which can not be constrained using the vertex 
operator argument for the Cartan generators. Possible terms which 
violate the shift symmetry can be viewed as part of the 
corrections to the K\"ahler potential due to boundary changing 
operators. They are suppressed by an extra power of the Planck mass and do 
therefore not affect the phenomenological analysis of the previous section.

On top of that, the shift symmetry can be broken at tree-level by couplings to
fields charged under the SM.  The latter, however, are innocuous
for our application because fields charged under the SM gauge group
will have neither a bosonic VEV nor, generically, a non-trivial F-term. Thus
such terms do not affect the structure of the Higgs mass matrix resulting from
the tree-level K\"ahler potential.

The symmetry breaking (\ref{ad-dec}) can be  implemented  by quotienting the
compactification space $X_3$ by a discrete symmetry group $\mathbb G$ (which
restricts to a symmetry of the 3-cycle $\Sigma$ wrapped by the D6-brane) and
suitably embedding the action of $\mathbb G$  into $G$.  In the case of a
freely-acting symmetry group this is just the Hosotani-Witten mechanism of
switching on discrete Wilson lines.  Massless matter descending from $\Phi$ is
given by the zero modes of the Laplace operator acting on 1-forms on $\Sigma$
twisted by the flat connection corresponding to the Wilson line. This, however,
leaves the components of $\Phi$ in the adjoint of $H$ massless because these
correspond to the zero modes of the untwisted Laplace operator, which are
unaffected. (Put differently, freely acting quotients do not reduce the
fundamental group and so $b_1(\Sigma)$ cannot decrease.)

What remains possible is to quotient $X_3$ by a discrete symmetry group
$\mathbb G$ that restricts to a non-freely acting symmetry of  $\Sigma$. This
corresponds to an orbifold $X/\mathbb G$  with non-trivial fixed-points, even
though the covering space $X_3$ is not necessarily toroidal. Nonetheless the
string theory on $X_3/\mathbb G$ is well-defined as long as $\mathbb{G}$ is
embedded into the holonomy group $SU(3)$. The cycle $\Sigma$ wrapped by the
D6-brane must pass through some of the fixed-points so that the brane is mapped
to itself and the brane spectrum is projected.  For a concise summary of this
projection technique in the context of toroidal Type IIB branes see e.g.
\cite{Aldazabal:1998mr} and references therein, and the same methods apply to
Type IIA branes. 

Let us now comment on the situation for gauge theories on 7-branes in Type IIB
orientfolds or F-theory.  As reviewed at the end of section \ref{sec_ShiftOP}, 
there are two types of open string
moduli of a 7-brane wrapping a holomorphic divisor $S$. The complex Wilson line
moduli counted by elements in $H^1(S)$ exhibit a shift symmetry, while for the geometric deformations given by\footnote{Here $K_S$ is the canonical bundle of $S$ and ${\cal
O}$ the trivial bundle.}  $H^0(S, K_S)
\simeq H^2(S, {\cal O})^*$ this symmetry will generically be broken
in the manner discussed above.  

If we start with a gauge group $G$, both types of fields give rise to chiral
multiplets in the adjoint representation, which descend to matter in suitable
representations upon breaking $G \rightarrow H \times F$. In addition to
symmetry breaking via orbifolds, Type IIB models offer the possibility of
considering non-flat gauge connections, i.e. non-trivial gauge flux.  In the
presence of gauge flux, the divisor $S$ is endowed with a non-trivial
holomorphic vector bundle $L$, whose structure group $F$ is embedded into $G$.
For simplicity we can focus on line bundles by taking $F= U(1)$.  The massless
${\cal N}=1$ chiral superfields in representation $R_i$ with U(1) charge $q_i$
are given by the zero modes of the  twisted Laplace operator acting on one- and
two-forms. Concretely, the chiral bulk spectrum is given by
\bea
H^1(S, L^{q_i})  \oplus      H^2(S, L^{ - q_i} ).  
\eea
The chiral superfields in $\bar{R_i}$ are counted by
\bea
H^1(S, L^{- q_i})  \oplus      H^2(S, L^{ q_i} ) . 
\eea
See e.g. \cite{Beasley:2008dc,Blumenhagen:2008zz} for a recent discussion of
details of the matter spectrum on 7-branes.  It is tempting to ascribe to the
elements in $H^1(S, L^{q_i})$ a Wilson line shift symmetry. However, unlike in
the case of an orbifold, these states are \emph{not} directly related to a
``universal'' Wilson line in the adjoint of the underlying gauge group $G$,
which would be counted by $H^1(S, {\cal O})$.  It is therefore not obvious that
the shift symmetry argument for such an underlying Wilson line implies a
K\"ahler potential of the form (\ref{Kahler4}).\footnote{
Indeed, 
the CFT argument for the shift symmetry only holds for flat Wilson lines. Note,
however, that the vector bundles on a B-type brane are holomorphic, i.e. their
curvature satisfies $F^{2,0} = 0 = F^{0,2}$. While this implies that 
curvature is non-zero in certain directions, one may hope that others are 
still protected from non-derivative couplings.}

Some helpful light can be shed on this question by comparison with heterotic
orbifolds.  The blow-up of such orbifolds corresponds to a smooth heterotic
compactification with non-trivial vector bundles \cite{Honecker:2006qz}.  Both
the twisted and the untwisted matter maps to different elements of the same
cohomology group on the resolved space (see also the discussion in the appendix
of \cite{Tatar:2008zj}). This suggests that one combination of these fields
\emph{can} be viewed as the analogue of orbifold bulk matter. We reiterate,
however, that the resolution process might significantly correct the shift
symmetric K\"ahler potential due to higher-order terms involving the blow-up
modes. We leave it for future work to study the shift symmetries of bulk matter
in presence of vector bundles.

Models with bulk matter fields have recently been re-addressed also in the
context of F-theory GUT model building. Compatibility of a bulk Higgs field
with the desired Yukawa structure often requires non-perturbative effects in
form of exceptional singularity enhancements as considered recently in
\cite{Donagi:2011dv}.  The construction of concrete setups which in particular
accomodate the SM Yukawa couplings is an exciting challenge which 
we plan to address in the near future.

\section*{Acknowledgments}
We would like to thank L. Anderson, R. Blumenhagen, T. Gherghetta, 
J. J\"ackel, $\mbox{H.~Jockers}$, 
D. L\"ust, M. Luty and F. Marchesano for discussions. AH thanks the 
School of Physics of the University of Melbourne for hospitality. TW thanks 
the IFT at the Universidad Autonoma, Madrid, and the Simons Center for 
Geometry and Physics, SUNY Stony Brook, for hospitality.

\end{document}